# A Decade in a Systematic Review: The Evolution and Impact of Cell Painting

*Srijit Seal*[1,2]‡*, *Maria-Anna Trapotsi*[3]‡*, Ola Spjuth[4], Shantanu Singh[3], Jordi Carreras-Puigvert[4], Nigel Greene[5], Andreas Bender[2], Anne E. Carpenter[1]*

[1]Broad Institute of MIT and Harvard, Cambridge, Massachusetts, United States

[2]Yusuf Hamied Department of Chemistry, University of Cambridge, Lensfield Road, CB2 1EW, Cambridge, United Kingdom

[3]Imaging and Data Analytics, Clinical Pharmacology & Safety Sciences, R&D, AstraZeneca, 1 Francis Crick Avenue, Cambridge, CB2 0AA, United Kingdom

[4]Department of Pharmaceutical Biosciences and Science for Life Laboratory, Uppsala University, Box 591, SE-75124, Uppsala, Sweden

[5]Imaging and Data Analytics, Clinical Pharmacology & Safety Sciences, R&D, AstraZeneca, 35 Gatehouse Drive, Waltham, MA 02451, USA

‡ These authors contributed equally.

*seal@broadinstitute.org

*marianna.trapotsi1@astrazeneca.com

*anne@broadinstitute.org

Machine Learning, Cell Painting, Drug Discovery, High Content Screening, Toxicity, Mechanism of Action

**ABSTRACT**

High-content image-based assays have fueled significant discoveries in the life sciences in the past decade (2013-2023), including novel insights into disease etiology, mechanism of action, new therapeutics, and toxicology predictions. Here, we systematically review the substantial methodological advancements and applications of Cell Painting. Advancements include improvements in the Cell Painting protocol, assay adaptations for different types of perturbations and applications, and improved methodologies for feature extraction, quality control, and batch effect correction. Moreover, machine learning methods recently surpassed classical approaches in their ability to extract biologically useful information from Cell Painting images. Cell Painting data have been used alone or in combination with other -omics data to decipher the mechanism of action of a compound, its toxicity profile, and many other biological effects. Overall, key methodological advances have expanded Cell Painting's ability to capture cellular responses to various perturbations. Future advances will likely lie in advancing computational and experimental techniques, developing new publicly available datasets, and integrating them with other high-content data types.

**INTRODUCTION**

Phenotypic drug discovery (PDD) identifies compounds based on their ability to alter a given disease phenotype. PDD has been a critical component of therapeutic development[1], having evolved from screening a few compounds in animals to testing millions in cell models. Although target-based drug discovery (TDD) bore much fruit throughout the 20th century[2], scientific advancements such as development of gene editing tools, organoids, and imaging assay technologies, as well as the increasing awareness that understanding the exact molecular target of a compound, as required in TDD, is not always a prerequisite for effective and safe therapeutic discovery. This has resulted in the resurgence of phenotypic screening approaches.[1] This is evident from the fact that around 7–18% of FDA-approved drugs do not have a defined molecular target[3], and several drugs have been found not to work via their purported target, as highlighted in a recent analysis on anti-cancer drugs that therapeutically acted through off-target effects.[4] Therefore, phenotypic strategies have gained favor because they also allow compounds to be explored in a target-agnostic manner (using hypothesis-free assays).[5]

Although target-based strategies (using hypothesis-based assays) have been valuable in specific therapeutic areas, they have limitations, particularly when targets are complex or considered undruggable, and when the disease of interest is polygenic.[6] Swinney and Anthony highlighted that, out of the 50 small molecules discovered as first-in-class small molecule drugs with new molecular mechanisms of action (MoA) between 1999 and 2008, target-based strategies discovered 34% of these (17 small molecules) while phenotypic strategies discovered 56% (28 small molecules) with the remaining 10% (5 small molecules) being synthetic versions of natural substances.[7] However, this does not mean that target-based drug discovery is inherently inferior; phenotypic drug discovery is more likely to succeed when understanding the most desirable mechanisms of action is prioritized.[1]

Among phenotypic screening strategies, High Content Screening (HCS) technologies have proven to be both effective and efficient, allowing multiple parameters to be measured at the single-cell level simultaneously.[1] HCS enables cellular complexity and heterogeneity to be captured in response to a range of perturbations, such as genetic modifications, environmental stressors, or small molecule treatments. At the core of these technologies is cellular morphology—the visual appearance of cells, usually stained for cell structures or biomarkers—which is intricately linked to cell physiology, health, and function.[8] HCS has a broad spectrum of applications in biological and drug discovery research. For example, CRISPR-Cas9, siRNA, and cDNA screens are used to identify genes and proteins involved in specific pathways and processes and are also applied in academia and pharmaceutical companies for target identification.[9,10] HCS is also used in drug discovery to screen for novel compounds, and to better understand the biological effects of compounds. For example, compounds identified through traditional screening could be profiled further using phenotypic assays to investigate selectivity and toxicity, such as in-vitro micronuclei formation assays to identify compounds that could potentially damage DNA.[11,12] These assays investigate specific endpoint(s) and require a careful experimental design; many parameters need to be considered, such as the selection of cell model or cell line, growth conditions, biomarkers, dyes, or antibodies.[11]

A major development came on the scene in 2004, when Perlman et al. demonstrated that instead of tailoring an image-based assay to a particular biology of interest, images might be used in an relatively unbiased way (besides choice of experimental conditions) to group drug treatments based on having similar impacts on cell morphology.[13] This finding, in combination with other developments such as the building momentum of transcriptional profiling[14] and the technology push to generate more biological data, resulted in the launching of the field of image-based profiling and the development of image assays aiming to maximize information content. The most popular among them is the Cell Painting assay, first described by Gisladóttir et al. in 2013[15]. The Cell Painting assay generates a holistic "painting" of the cell that reflects its phenotypic state and cellular responses to perturbations. The Cell Painting assay (Figure 1a) involves staining cells with a combination of fluorescent dyes, each labeling distinct cellular components or organelles. The most widely used dyes to perform this assay are Hoechst 33342 (DNA), concanavalin A (endoplasmic reticulum), SYTO 14 (nucleoli and cytoplasmic RNA), phalloidin (f-actin) and what germ agglutinin (WGA) (Golgi apparatus and plasma membrane), and Mito Tracker Deep Red (mitochondria).[16] The Cell Painting assay was designed to be easy and inexpensive to implement in any high throughput screening facility, relying solely on dyes rather than antibodies, which can be more costly and involve multiple labor-intensive steps. This multiplex staining approach is followed by processing with automated imaging pipelines (such as CellProfiler[17]) that can then extract morphological profiles and standardize them against reference and control compounds (Figure 1b). This yields a high-dimensional dataset for each cell and captures over a thousand morphological features (including measures of size, shape, texture and intensity, among many others). This is then followed by normalization/pre-processing, and batch effect corrections. The morphological profiles can then be used for further downstream analysis by distinguishing perturbations based on their phenotypic responses (Figure 1c). Cell Painting profiles can also be used with unsupervised machine learning models, for example, to detect clusters with similar MoAs, or with supervised machine learning models (for example, to predict the MoA or toxicity of new compounds).

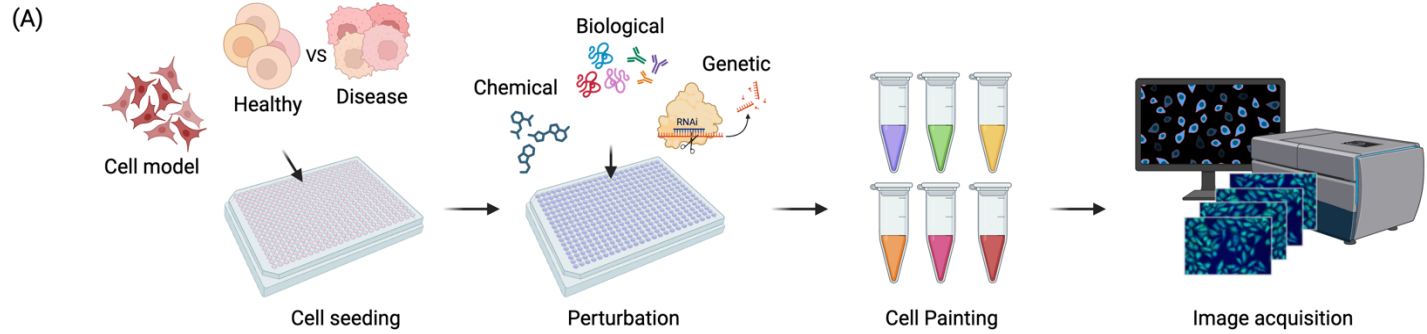
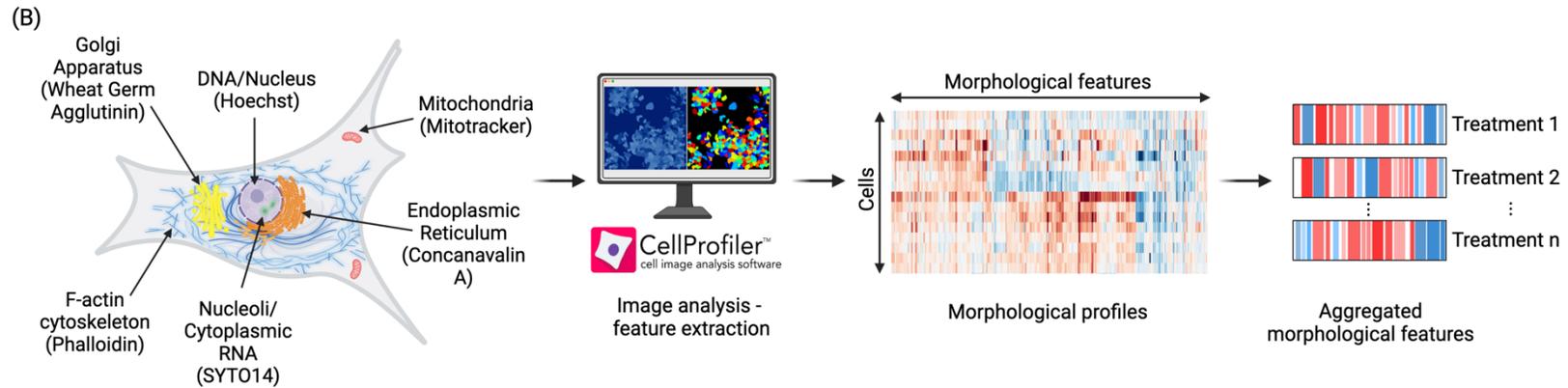
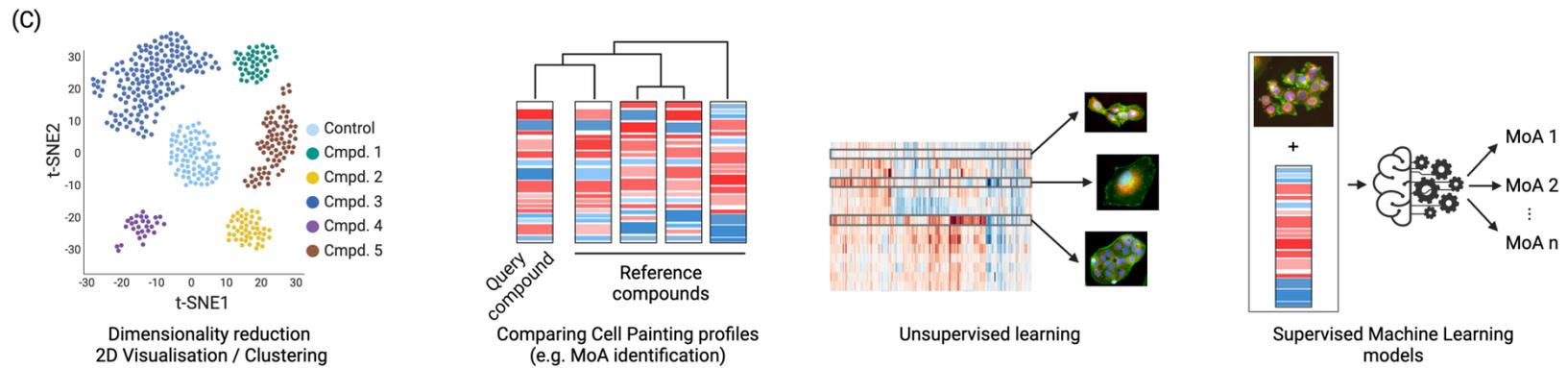

**Figure 1.** Morphological profiling using the Cell Painting assay. a) Schematic representation of Cell Painting assay; cells are incubated and perturbed and a set of six stains is applied. Images are then obtained by automated microscopy followed by nucleus and cell body segmentation. b) Appropriate software or deep learning-based methods are applied to measure or calculate morphological features from the images. c) After feature pre-processing, downstream analysis is performed. This includes a variety of methods, including supervised and unsupervised machine learning, to better elucidate the biological effects of a compound, such as its MoA or safety profile.

The Cell Painting assay has seen increasing adoption in academic and industry research and here we aim to comprehensively examine the advancements and impacts of Cell Painting in drug discovery and related areas over the past decade (2013–2023), following a systematic review format. We explore the methodological advancements that have improved the robustness of the assay and discuss how Cell Painting has deepened understanding of disease processes and shaped therapeutic discovery. Importantly, we discuss the integration of Cell Painting with machine learning and other -omics data. Moreover, we explore the role of Cell Painting in predictive toxicology and its significance in improving the safety and efficacy of drugs. Overall, we provide a comprehensive perspective on the potential of the Cell Painting assay and its impact in drug discovery.

## RESULTS

**Study Selection**

For our systematic review, we retrieved 340 articles from three sources: PubMed[18], Scopus[19], and ScienceDirect[20] (accessed June 2023). Among these, 207 duplicates were removed, and 41 review articles were further removed during the screening process. A total of 92 articles were eligible for a full-text analysis and of these 21 studies were further removed (18 irrelevant and 3 poster/thesis/news articles). To augment this, a manual search contributed an additional 18 studies, acknowledging the rapidly evolving nature of this field. Some of these studies were published after the initial cut-off date (i.e., after June 2023) but were still included during the review process due to their significant contribution to Cell Painting research. Overall, this yielded 89 studies for systematic review, as shown in the Preferred Reporting Items for Systematic Reviews and Meta Analyses (PRISMA) flow chart in Figure 2 (for a list of the 89 studies, see Supplementary Table S1 ; excluded studies in Supplementary Table S2 released at https://github.com/srijitseal/CellPainting_SystematicReview).

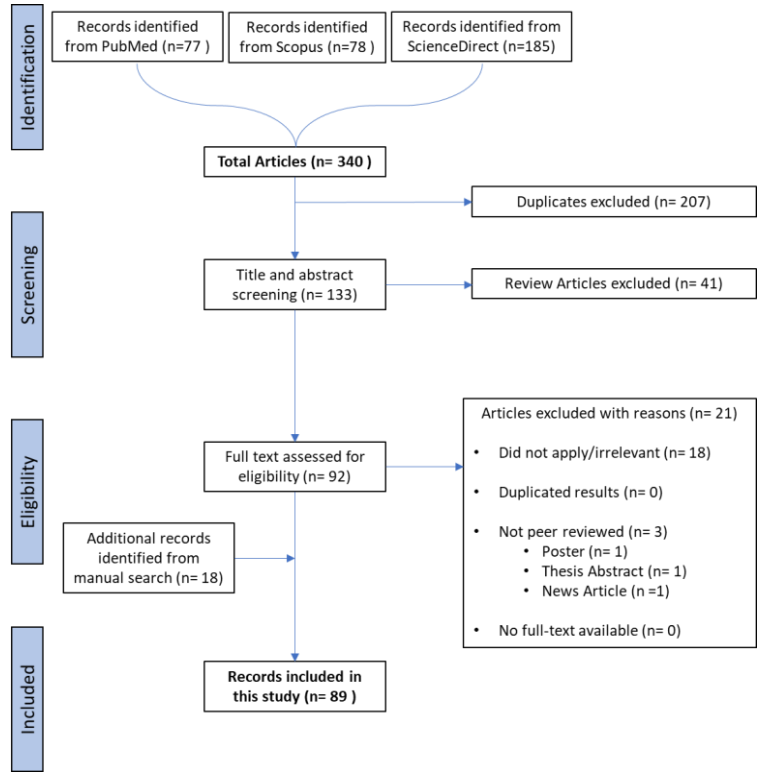

**Figure 2.** the Preferred Reporting Items for Systematic Reviews and Meta Analyses (PRISMA) flow chart diagram on the selection of the 89 studies included in this systematic review. Records from manual search included some articles published the June 2023 cut-off date.

**Extracted Data**

The extracted data included authors, year of publication, keywords, and the journal where the article was published. We manually identified the research question and the major outcome and categorized the results into major and sub-categories (where possible). Figure 3a shows the number of publications per year. Imaging-based profiling assays such as the Cell Painting assay are increasingly being used, as the majority of the studies were published within the last three years (2021–2023). This also indicates that, as with other new technologies such as transcriptomics data, it takes about a decade for scientists to make use of it on a larger scale. Figure 3b shows the number of publications from each journal in this study. The majority of papers were published in SLAS Discovery, which indicates that the assay was quickly accepted by the drug profiling and screening community. Computational journals, for example related to cheminformatics, and bioArxiv preprints (which maybe currently in review process at other journals) have also been a preferred dissemination route for findings from Cell Painting datasets. Journals for the fields of chemical biology and toxicology are among the other top publication choices, where Cell Painting data has been used for drug discovery.

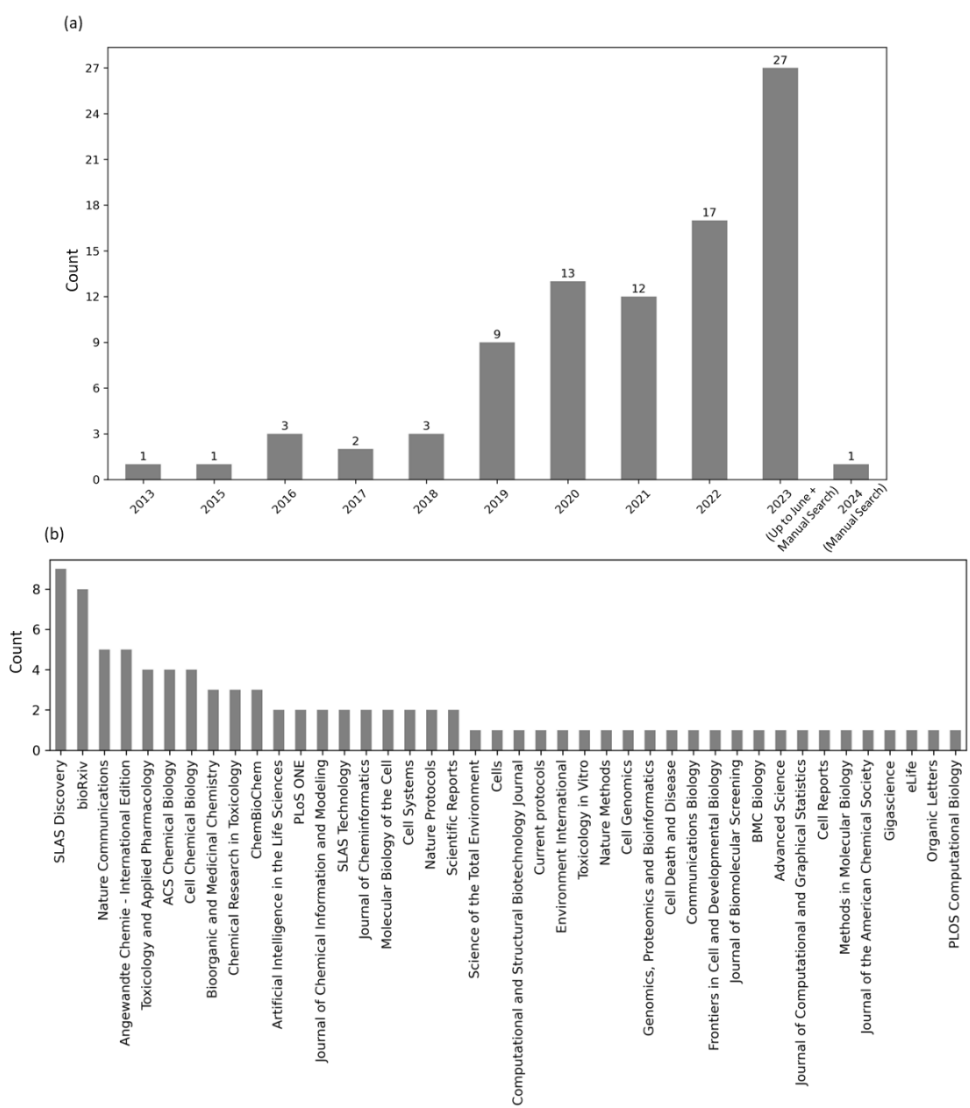

**Figure 3.** (a) The growth in publications reviewed in this systematic review between 2013 and 2023 and (b) the journals where the publications were published.

## DISCUSSION

### 1. Assay Development

The Cell Painting protocol was first developed by Gustafsdottir et al. in 2013. It was designed to be a low-cost single assay capable of capturing many biologically relevant phenotypes with high throughput.[15] As shown in Figure 1b, six stains were selected and imaged in five channels to reveal morphological changes for eight cellular components or organelles. Gustafsdottir's publication did not name the assay; however, an updated protocol (v2) published in 2016 by Bray et al. established the moniker "Cell Painting", while also making minor adjustments such as stain concentrations.[16] A recent effort optimized the assay's stability and reproducibility, culminating in Cell Painting v3 in 2022.[21] To create the updated protocol, the JUMP-CP (Joint Undertaking for Morphological Profiling – Cell Painting) Consortium, led by the Broad Institute, used a positive control plate of 90 compounds covering 47 diverse mechanisms of action to, for the first time, *quantitatively* optimized staining reagents, as well as experiment and imaging conditions.[21] These included reducing steps (such as no media removal before MitoTracker), decreasing dye concentrations to save reagent costs (such as the reduction of Phalloidin), and increasing SYTO 14 concentration to enhance signal-to-noise ratios. Another study included conditions such as time point and image acquisition conditions.[22]

#### 1.1 Cell Line Selection

For image-based assays, flat cells that rarely overlap are best – most cell lines meet this criteria. In general, dozens of cell lines have been used and have performed well for Cell Painting experiments, and thus the selection often depends on the purpose of the experiment. For example, A549 (lung adenocarcinoma) and U2OS (osteosarcoma) were considered for the JUMP-CP data, and both could have been suitable but U2OS was selected due to fewer restrictions related to sharing with partners.[21,22] In a recent study, the selection of optimal cell lines for high content screening was investigated because different cell lines can result in different sensitivities to detect different MoAs.[23] In total 3,214 small molecule annotated (i.e. containing information about their putative target and MoA) compounds (including FDA approved drugs) were profiled with Cell Painting on six different cell lines. These were the A549, OVCAR4, DU145, 786-O, HEPG2 and a noncancer patient-derived fibroblast cell line (FB). The cell lines were ranked based on their ability to infer compound activity termed "phenoactivity" (also "phenotypic activity") of the compound and MoA termed "phenosimilarity" (also "phenotypic consistency") of the mechanism groups). Here, compound activity refers to the observable effects a compound has on biological systems, such as inhibiting or promoting cellular processes in assays, while MoA describes the specific biochemical interaction through which a compound produces its effect, such as binding to a receptor or inhibiting an enzyme. Results showed that the best performing cell line in terms of detecting phenotypic activity may have poor sensitivity for compounds with the same MoA and *vice versa*. This discrepancy could be attributed to the diverse genetic landscapes of different cell lines, which may influence the expression of targets and the cellular pathways involved. For example, HEPG2 cell line's tendency to grow in highly compact colonies was identified as a factor impeding its ability to produce clear phenotypic distinctions between compound-treated and control groups. This growth pattern complicates the analysis of cellular organelles like mitochondria and actin—critical markers in the study—since alterations in these densely packed cells become harder to detect.

Another study showed that Cell Painting sample preparation protocol was effective without any cell line-specific adjustment for 16 reference chemical across six biologically diverse and morphologically distinct human-derived cell lines (U-2 OS, MCF7, HepG2, A549, HTB-9 and ARPE-19) that originate from various tissues and exhibiting distinct growth patterns and cellular characteristics.[24] It was only necessary to optimize image acquisition and analysis parameters to account for differences in the size and 3D shape of each cell line when cultured in monolayers. The maximum ranges for the *in vitro* point-of-departure (concentration at which chemicals begin to perturb cellular biology) differed by 1.05 $\log_{10}$ units across cell types, and the smallest ranges differed by 0.35 $\log_{10}$ units. Most of the chemicals tested showed a pronounced phenotypic effect across all cell lines, often below cytotoxic and cytostatic concentrations. However, for all chemicals except one, the most sensitive features were different in each cell line. This indicates that, even though the concentration at which chemicals alter cellular morphology is consistent across cell types, the way these effects manifest depends on the biological context of the cells. Over the past decade, the basic Cell Painting protocol has been used on dozens of additional cell lines without adjustment based on our survey of literature and personal communications (Personal Communications, March 15, 2024, Dr. Anne Carpenter).

**1.2 Adaptations of the Cell Painting Assay**

Adaptations of the Cell Painting assay have begun to appear, which replace some of the original dyes with alternative fluorescent dyes to increase the spectral range and facilitate delineation of other cellular compartments and structures.[25] One such assay, the HighVia assay, uses direct single-cell analysis to evaluate mechanisms of cell death (such as apoptosis and necrosis), as well as determine *in vitro* IC50, a measure of potency, for compound perturbations.[25] Another study replaced one dye (MitoTracker) with an antibody against a viral protein introducing the possibility of multiplexing Cell Painting with specific targets.[26]

Another adaptation of Cell Painting is LipocyteProfiler, which incorporates BODIPY to mark lipid droplets - this metabolic disease-oriented phenotypic profiling system is used for lipid-accumulating cells.[27]

**1.3 Improvements in the Choice of Perturbations**

In addition to the improvements made to the assay protocol, there are also recommendations that can help researchers tailor the screen's conditions and parameters with respect to downstream analysis. In addition to compound perturbation, Singh et al. first explored RNAi-induced knockdown using the Cell Painting assay in 2015.[28] They showed that the morphological signatures are highly sensitive and reproducible but there were off-target 'seed' effects of RNA interference reagents that dominated the signatures. These 'seed' effects occur when a short region of the RNAi molecule, known as the 'seed' sequence, binds non-specifically to multiple messenger RNAs. More successful technologies include open reading frame (ORF) constructs that allow for gene/protein overexpression[29] and CRISPR knockout to deplete expression.[30] One challenge with a target agnostic assay, such as Cell Painting, is that compounds active in the assay can act by on- and/or off-target effects, which in turn results in the difficult interpretation of a given bioactivity.[31] Hence, one practical solution is to include known reference compounds. Most recently, Chandrasekaran et al. and Jamali et al. introduced various sets of recommended control and landmark perturbations — including two compound plates and ORF and CRISPR perturbations plates.[22,32] Dahlin et al. generated a set of Cell Painting and cellular health profiles for 218 prototypical cytotoxic and prototypical 'nuisance' compounds in U2OS cells in a concentration-response format.[31] 'Nuisance' compounds, in this context, are substances that frequently show up as hits in screening assays but are ultimately considered undesirable because their effects are often nonspecific, artifactual, or due to properties that interfere with the assay rather than a specific biological activity of interest. The exploration and analysis of this dataset highlighted relationships between different types of cellular injury with Cell Painting activity. Robust Cell Painting phenotypes were observed with compound-mediated cellular damage (e.g. tubulin poisons). This reference dataset thus serves as a valuable resource for comparing and triaging novel compounds, especially when they exhibit phenotypes similar to reference compounds with undesirable MoAs.

**1.4 Development of Microscopy Imaging**

Although high-throughput imaging platforms have advanced over the past decade resulting in improved speed and resolution, Jamali et al. found that various microscope imaging systems performed similarly and changing acquisition settings only minimally affected Cell Painting profile strengths.[32] Key setting alterations that improved morphological signatures included decreasing magnification, which increases the number of cells imaged. Jamali et al. concluded with a general set of recommendations for Cell Painting, applicable to several microscopes, suggesting that cells should be imaged at 20x magnification across four to nine sites (fields of view), or approximately 2,500 cells per well for the cell types considered in the study.

**2.1 Extraction of Morphological Features from Fluorescent Images**

Cell Painting images are often analyzed using software to extract morphological features, enabling the precise segmentation of cellular and subcellular structures. The open-source CellProfiler[17] software is one example; however, other solutions are also used, including proprietary ones (for a detailed review see Smith et al.[33]). Some alternative analytical approaches use deep learning (such as the recently published DeepProfiler[34]), where models are trained to recognize features directly from raw images to optimize Cell Painting profiles, increasing their sensitivity and robustness, and in some cases skipping the single-cell segmentation step. Steps to further process the Cell Painting data, from morphological feature extraction to profile normalization to batch effect correction discussed in Section 3.1 and 3.2, are also continuously improving.

**2.2 Extraction of Morphological Features from Label Free Brightfield images**

Some researchers have investigated 'label-free' assays replacing fluorescent images with brightfield (BF), which can be captured for living cells over time. Even for fixed cells, fluorescence imaging is typically more time-consuming, expensive and labor intensive compared to BF imaging. While BF imaging does not yield a clear contrast of the cellular compartments, the use of deep learning methods could potentially augment the information available in BF images. In one study, deep learning models were used to predict five Cell Painting fluorescent channel images from brightfield images and CellProfiler features were calculated from the predicted images and the ground truth images.[35] The models were trained on approximately 3,000 images (using one field of view per well from 17 batches) and then tested with 273 images. The predicted images achieved a mean Pearson Correlation of 0.84 with the ground truth at the pixel level, and the authors further calculated the Pearson correlation of CellProfiler features from the ground truth images and the predicted images from BF. Although many morphological features extracted from the generated images showed substantial correlation with those from the ground truth images (>0.6 correlation) and 30 features showed a correlation greater than 0.8, the features from the AGP and Mito channels were more challenging to predict, likely due to the small and subtle cellular substructures they contain. They additionally performed a downstream analysis and investigated the ability of models in predicting compounds similar to the positive controls, which resulted in a sensitivity of 62.5% and specificity of 98.0% and one reason why sensitivity was low could be due to potential useful information missed from the Mito and AGP channels.

Another study compared the abilities of deep learning-based features to predict ten MoA classes.[36] The features were trained using either brightfield images (BF) or fluorescence images and were additionally compared to the benchmarked CellProfiler features from fluorescent images. The dataset consisted of 231 compounds with 10 MoAs tested at 10uM on U2OS cells. Models trained with features from BF images, fluorescent images and CellProfiler features predicting the class labels for 11 classes (10 MoAs and the DMSO control class) showed comparable results.[37] Using activation maps, they further determined which areas in the images were most activated and this revealed that the models focused on different cellular features depending on the image type used for training. For example, when predicting the MoA for the compound, 4SC-202, the models had an accuracy of 0.89, 0.04 and 0.29 when using BF, fluorescent images and CellProfiler features respectively. The better performance of BF might be due to the fact that in the BF heatmap, there was a strong activation for the small vesicles that are visible in the BF images but these are not stained in the Cell Painting protocol (the FL heatmap focuses is on the full cell body). Despite the limited number and range of MoAs tested, this study suggests that deep learning applied to brightfield images holds great promise to augment or replace fluorescent stains in Cell Painting assays in the future. This change could lead to significant time and cost savings in future experiments by leveraging the vast imaging resources already available from brightfield microscopy, which is more widely used than fluorescence imaging. Early reports from the techbio company Recursion indicate a transition from fluorescent to brightfield imaging.[38]

**3.1 Feature Selection for Cell Painting Profiles**

Not all morphological features extracted from cell images, using e.g. CellProfiler, are informative. For a given task or even for a general representation of cell phenotype, feature selection methods are generally used to filter them and are available from virtually all data analysis libraries (e.g. scikit learn).[39] Pycytominer, a software package designed for analyzing Cell Painting data, incorporates feature selection methods that reduce redundancy and increase informativeness of features[40]. Other approaches, such as AutoML (automated machine learning) in Siegismund et al, enable the most informative features from Cell Painting datasets to be identified faster.[41] The AutoML approach presented in Siegismund et al showed that a subset of only 20–30 features was sufficient to represent the most relevant information from the morphological signature and successfully differentiate between the control class and perturbations; although this will likely depend on the endpoint being classified and the amount of data and the diversity of phenotypes in the profiled dataset.[41]

**3.2 Normalization and Batch Corrections for Cell Painting Profiles**

The importance of experimental design in Cell Painting assays should not be underestimated because it can substantially impact the efficacy of normalization methods.[42] For example, it is important to minimize confounding factors related to batch effect variability within the morphological features. For example, Janosch et al. explored the selection of unbiased features solely using dimensionality reduction methods on images from negative controls.[43] For Cell Painting datasets, the Pycytominer tool normalizes data for individual plates, either by using all wells or solely the negative control wells (using the RobustMAD method).[40] Pycytominer also implements the sphering transformation (also termed as "whitening") of morphological signatures which can be viewed as a multivariate standardization strategy.[30,44] Sphering the profiles increased the percent replicating score – a measure of reproducibility of replicates of each sample – from 18–37% to 83–84% (for compounds at 10uM), although these results have not been consistently high across studies, and may be confounded by plate layout effects.[44]

As part of the JUMP Cell Painting Consortium, Arevalo et al. conducted a comprehensive analysis of seven batch effect correction methods selected from a single-cell mRNA profiling benchmark study.[45] They used qualitative visualizations in combination with 10 metrics to assess performance on image-based profiles, focusing on batch effect reduction and preservation of biological signals. These methods were applied to JUMP Cell Painting Consortium data for five scenarios of increasing complexity: batches from within and between different laboratories, within and between different imaging equipment, and with low and high numbers of replicates.

Recent studies have begun to explore the potential of deep learning models for batch correction, aiming to separate noise from true biological signals in Cell Painting data. Yang et al. investigated a mean teacher-based model called DeepNoise, which was tested on the RxRx1 dataset consisting of 125,510 fluorescent microscopy images from Recursion for the CellSignal competition.[46] The study found that DeepNoise effectively distinguished biological phenotypes from technical variations, achieving a multiclass accuracy of 99.60% compared to 74.58% using plate-based normalization. However, the evaluation of the method had limitations due to the inaccessibility of the test dataset labels, which prevented extensive comparisons with other models and further analysis of predictions on each of the four cell types, as well as the calculation of additional metrics such as specificity and sensitivity. Despite these limitations, the study indicates that deep learning methods may be more effective at learning batch-effect patterns in Cell Painting datasets compared to standard normalization approaches.

## 4. Publicly Available Datasets

Over the past decade, Cell Painting has been used to screen chemical compounds and genetic perturbation libraries to facilitate the extraction of phenotypic information from cells. The resulting datasets have provided invaluable insights into multiple areas of the life sciences. Image sets of varying sizes have been made available publicly in different locations, including the Image Data Resource[47], the Broad Bioimage Benchmark Collection[48], and university websites. Recently, the Cell Painting Gallery was launched to provide a central location for datasets of interest, hosted through Amazon Web Service's Registry of Open Data.[49] Currently, three large Cell Painting datasets are publicly available for compound and/or genetic perturbations (Table 2), which contain many thousands of perturbations and are being used to study morphology signatures.

Table 2: Large Publicly Available Cell Painting Datasets covering Compound or Genetic Perturbations

| Dataset | Release Date | Type of Perturbations | Cell Line | Number of unique perturbations (compounds or genetic) | References |
| --- | --- | --- | --- | --- | --- |

| Bray et al. Cell Painting Dataset | January 2017 | Compounds | U2OS | 30,616 bioactive compounds | [50] |
| --- | --- | --- | --- | --- | --- |
| Recursion RxRx3 dataset | January 2023 | CRISPR/Cas9-mediated gene knockouts and compounds with dose-response | HUVEC | 17,063 CRISPR/Cas9-mediated gene knockouts (most anonymized) and 1,674 compounds at 8 concentrations each | [51] |
| JUMP-CP dataset | March 2023 | Over-expression of genes, knockout of genes using CRISPR-Cas9, and compounds | U2OS | Over-expression of 12,602 genes, knockout of 7,975 genes using CRISPR-Cas9, 116,750 unique compounds | [22] |

Wawer at al. from the Broad Institute released the first large public Cell Painting dataset in 2014; the same data was further refined and re-released in Bray et al. 2017, covering over 30,616 compounds in human U2OS cells.[50,52] More recently, Recursion Pharmaceuticals released their RxRx3 dataset, their largest to date.[51] Rxrx3 consists of approximately 2.2 million multi-channel microscopy images, representing over 1,674 unique compounds and 17,000 genes profiled in the HUVEC cell line. They gathered this data through their efforts to develop treatments for various diseases; however, there is one caveat: most genes are anonymized to protect their business interests. The JUMP-Cell Painting Consortium recently released a large-scale dataset involving the U2OS cell line, perturbed with over 116,750 different molecules, 12,602 gene overexpression reagents, and 7,975 gene knockouts using CRISPR-Cas9.[22] A comparison of the chemical space of compounds from these three large datasets is shown in Figure 4. Several studies have also released smaller datasets (derived from the Broad Institute's large public datasets) that provide different modalities of data along with Cell Painting. One recent study by Haghighi et al. provided a collection of four datasets covering 28,000 genetic and compound perturbations in both the Cell Painting and L1000 (for gene expression) assays.[53] Another study by Dafniet et al. released a chemogenomic library of 5,000 small molecules by integrating drug target-pathway-disease relationships with morphological profiles.[54] This dataset includes 2,473 chemical-target interactions covering diverse biological contexts that can be used in combination with Cell Painting morphological signatures accessible from the Broad Institute's datasets to facilitate target identification. Most recently, several pooled, genome-wide CRISPR perturbation screens' data were released, in which Cell Painting was adapted to an optical barcoding scheme.[55]

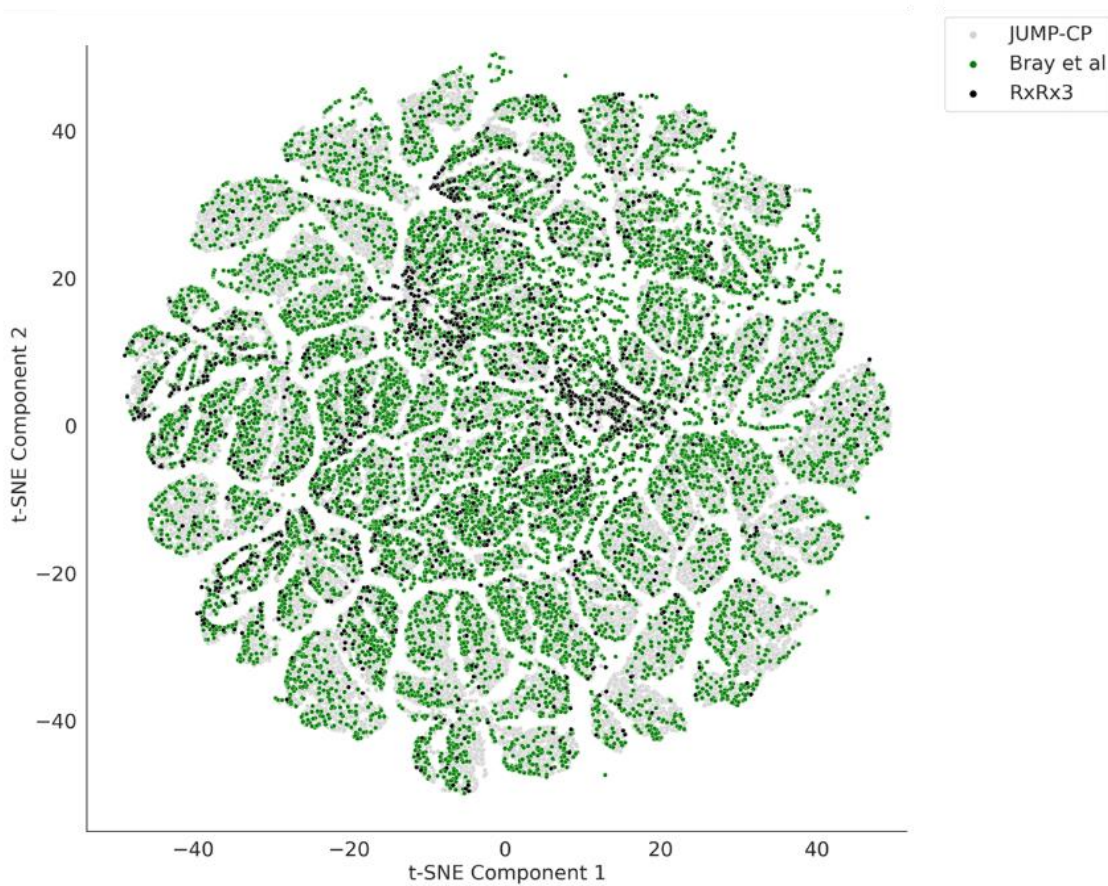

**Figure 4.** The physicochemical space of compounds in the three publicly available Cell Painting datasets. The physicochemical space was defined by using t-distributed Stochastic Neighbor Embedding (t-SNE) embeddings from six descriptors: molecular weight, topological polar surface area, number of rotatable bonds, hydrogen bond acceptors, hydrogen bond donors, and log P.

## 5. Applications of Cell Painting Data

As in most scientific domains, the processing and downstream analysis of Cell Painting data (Figure 1c), often using Machine Learning (ML) and statistical approaches, have enabled complex patterns in such data to be identified and accurate predictions to be made. ML models used in biological sciences research offer new frameworks for understanding biological systems through the lenses of information compression, qualitative intelligibility, and dependency relation modelling.[56] These algorithms are particularly well-explored for analyzing morphological profiles to predict the safety or toxicity of unknown compounds, both *in vitro* and *in vivo*, and to predict MOAs and targets.[57–59] Supervised methods are used when labeled datasets are available (that is, the "correct answer", or ground truth, is known for each sample), enabling the algorithms to be trained to predict specific outcomes and patterns from feature representations. Unsupervised methods are used to investigate the similarities among samples in the feature space itself without needing labelled data. Such clustering can be based on similarity, distance, or density metrics. Deep learning-based methods have also been applied to Cell Painting data, both to train predictive models and to calculate morphological features directly from raw images (rather than using classical algorithms). In the following sections we discuss the different applications of Cell Painting data, using ML and statistical approaches, to aid drug discovery.

**5.1 Predicting Mechanisms of Action**

The Cell Painting assay offers a comprehensive view of cellular responses to compound perturbations, enabling the identification of mechanisms of action (MoA) for compounds that induce morphological changes detectable by the assay. This identification process involves comparing the phenotype of a query compound to those of 'landmark compounds' with known MoAs. However, defining a compound's MoA is complex, as compounds can have multiple targets with varying affinities, and genes and proteins downstream of the direct targets may be altered differently in various cell types (e.g., expression levels, phosphorylation levels). As a result, binary labels for MoAs in datasets often oversimplify the reality (see Trapotsi et al. for more details).[60] Moreover, the resolution of MoAs that can be adequately described by any profiling assay, including Cell Painting, transcriptomics, or proteomics, is limited because these assays, individually or collectively, do not capture every possible cellular response. While they provide valuable insights into some MoAs, the applicability of Cell Painting data for each MoA must be established separately. It is important to note that 'Mechanism of Action' is a broad term, and studies in this section may discuss target-related MoA predictions, such as drug-target interactions, or biological process-related MoA predictions.

**5.1.1 Predicting Biological-process-related Mechanisms of Action**

Early studies in the field initially focused on exploring the efficacy of data derived from the Cell Painting assay to distinguish compounds' MoA under different experimental conditions, such as the number of channels and multiple cell lines. Cimini et al. analyzed 90 compounds and found that collected data are robust to changes in the measured channels, however datasets are small and specific phenotype(s) of interest may depend on compartments that are less critical for the compounds in the study.[21] Therefore, it would be interesting to reassess this with newer and larger publicly available datasets. A study by Rose et al. investigated the impact of using multiple fluorescent dyes for DNA, actin and tubulin, which partially overlap with the Cell Painting dyes, and 453 CellProfiler features were extracted in total to predict the MoAs of different compounds.[61] First, they investigated 38 drugs from the BBBC021 dataset which included 12 relatively distinguishable MoA labels (including actin disruptors, DNA damage, kinase inhibitors, and others), achieving up to 83% MoA prediction accuracy with three staining channels (DNA, actin, tubulin). Accuracy remained at 68% using the DNA channel alone. Rose et al. next determined whether using multiple cell lines with fewer markers could be more effective than using multiple dyes on a single cell line.[61] To do this, they used 10 cell lines with fewer markers to test 614 compounds, which targeted 113 gene products. By using an ensemble voting method (target within the top five predictions), they achieved an accuracy of 25% compared to a random classifier (accuracy 0.09%). They found that incorporating additional cell lines incrementally increased the overall prediction accuracy; however, as nine of the ten cell lines were biologically similar to each other, it would be useful to test whether including a more diverse set of cell lines could further improve prediction accuracies. A few years later, Cox et al screened 1,008 approved drugs and well characterized compounds (218 unique MoAs) at four concentrations in a high content screen against a panel of 15 reporter cell lines (three cell lineages, 12 organelle and pathway markers; grouped in five combinations).[62] Morphological profiles of these compounds (generated using PerkinElmer Acapella/Columbus software) were used for MoA prediction, and the treatments with active reference compounds were ranked based on their mechanisms of action. This ranking involved calculating the area under the receiver operating characteristic curve (AUC-ROC) for each MoA and a high AUC-ROC value (≥ 0.9) indicated that the MoA could be clearly distinguished from others. Results showed that 20 out of 83 MoAs were readily distinguished in the best single cell line. The number of distinguishable MoAs increased with the addition of each cell line but this effect quickly plateaued around 41 MoAs, that is, 41 out of 83 MoAs were readily distinguished across all 15 reporter cell lines.

The selection of channels and cell lines are not the only parameters that were investigated to improve MoA classification accuracy. For example, using the BBBC021 and the BBBC022 (U2OS cells treated with 1,600 known bioactive compounds) datasets, Janosch et al. explored the selection of unbiased features using images from negative controls, using only dimensionality reduction methods to improve compound MoA classifications.[43] They hypothesized that if a feature remains reproducible within all negative control wells, any significant changes would likely be due to a perturbation rather than a technical variation. They compared their method with a benchmarked dimensionality reduction method (L1-norm) and then classified MoAs using the reduced features. In both the BBBC021 and the BBBC022 datasets, their proposed method using binned, stable parameters produced the second-highest accuracy. Removal of the noisiest parameters improved MoA classification accuracies (from 17.66% to 20.19% for the BBBC022). However, their proposed method showed better generalization when they trained models without seeing a class, for example with 180 compounds as dopamine receptor antagonists, 118 compounds were misclassified by L1-norm, whereas 115 were misclassified by their proposed method. Therefore, this method can be used to further select features for downstream analysis and authors suggest that the method could be further improved by applying deep learning methodologies.

**5.1.2 Identifying Biological-process-related Mechanisms of Action using similarity based approaches**

Moreover, the similarity of query compounds to reference compounds has been extensively used as a strategy to better understand MoA. Svenningsen et al. used this strategy to investigate the MoA of 9-methylstreptimidone[63] They calculated the Pearson correlation between the Cell Painting profiles of 9-methylstreptimidone and reference compounds known for their effects on protein synthesis inhibition, DNA synthesis, and tubulin dynamics modulation. 9-methylstreptimidone had the highest similarity to cycloheximide, a known protein synthesis inhibitor. Further validation confirmed that 9-methylstreptimidone acts as dose-dependent protein synthesis inhibitor.

Other studies have used the same "guilt-by-association" strategy to assign the MoA, that is, by comparing test compounds with reference compounds to understand their biological processes. Autoquin, a previously uncharacterized autophagy inhibitor compound, was found to be similar to iron chelators and to induce S-phase arrest.[64,65] Similarly, diaminopyrimidine

DP68 was identified as a Sigma 1 (σ1) receptor antagonist.[66] Cell Painting-based similarity approaches have also led to the discovery of several unexplored small molecules that modulate microtubules[67], inhibitors of pyrimidine biosynthesis, and three novel scaffolds targeting dihydroorotate dehydrogenase (DHODH).[68] The most commonly detected MoAs in Cell Painting readouts include microtubule modulation[67,69], DNA damaging agents[30], mitochondrial membrane depolarisation[69–71], and inhibitors of the plasma membrane Na+ pump[69], among others. It is important to note that the 'guilt-by-association' strategy is limited by the number of compounds with known MoA annotations, making the development of comprehensive reference sets crucial for their success.

**5.1.3 Identifying Biological-process-related Mechanisms of Action using Deep Learning approaches**

Apart from using pre-defined classical Cell Painting features as described in most studies above, another way to use imaging data involves using convolutional neural networks (CNNs), a type of deep learning architecture, directly on images of cells generated from Cell Painting assays. CNNs are adept at identifying correlations between objects and extracting meaningful information from images and can therefore be used to classify cellular phenotypes (Figure 5).

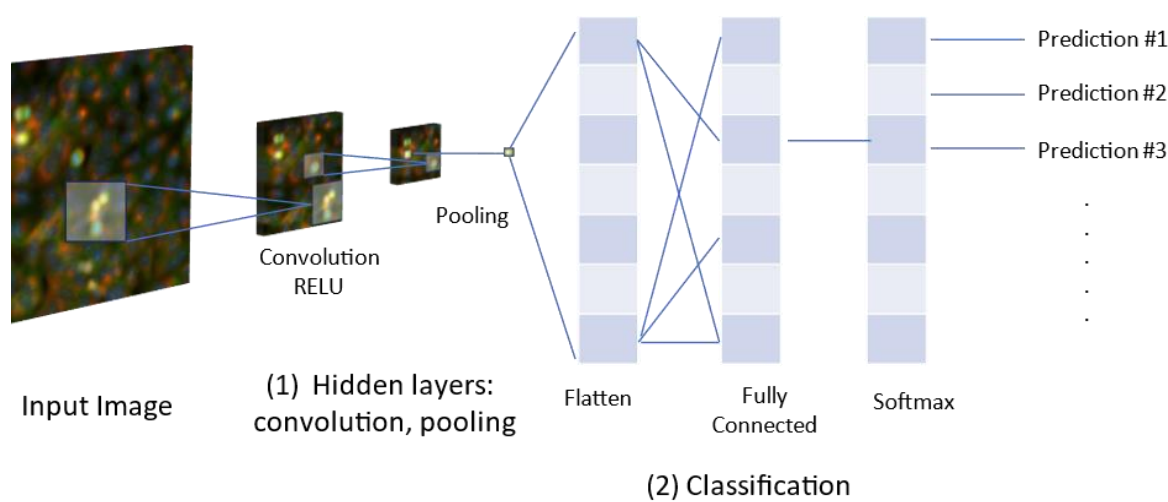

**Figure 5.** Summary of Convolutional Neural Network Analyses of Cell Painting Images. The input image consists of a matrix with pixel values. The convolution filters (smaller weight matrices) slide over the input image, detecting patterns such as edges, textures, and shapes, resulting in a feature map. An activation function (e.g., ReLU) is then applied element-wise, which introduces non-linearity into the model. Pooling then reduces the spatial dimensions of the feature maps (Step 1). Finally, the high-level features extracted from the image are flattened into a one-dimensional vector and arranged into fully connected layers to produce final classification scores for each category (Step 2).

One of the earliest studies by Durr et al. developed a CNN that classified single-cell phenotypes based on images generated from Cell Painting assays.[72] They trained CNNs to classify MoAs (including tubulin modulation, modulation of neuronal receptors, and cytotoxic MoAs) using approximately 40,000 single-cell images for 75 bioactive compounds and a DMSO negative control.[72] The CNN models misclassified 6.6% of all cells while models trained on numerical features from CellProfiler misclassified 8.9%, suggesting limited value in using CNNs compared to CellProfiler features. Applying CNNs to more challenging tasks, where traditional approaches do not perform well, will be useful to understand the potential added value of CNNs.

While the study by Durr et al. focused on using CNNs for directly classifying MoAs, another important aspect of deep learning in the context of Cell Painting is learning meaningful representations from the image data. These learned representations can capture important phenotypic features and confounding factors, which can then be used to improve downstream analyses, such as predicting MoAs or matching biological compounds. Several studies have explored various approaches to representation learning using deep learning in Cell Painting assays.

One such study by Moshkov et al. introduced a novel approach to learning representations from Cell Painting data. They trained a deep convolutional neural network architecture (known as EfficientNet) using only phenotypically strong compounds from three Cell Painting datasets. The features computed by EfficientNet accounted for both confounding factors and phenotypic features within the learned representation. This approach improved downstream analysis for matching biological compounds by 30% compared to using classical features. Studies such as Moshkov et al. highlight the potential of self-supervised learning strategies to learn meaningful representations that can enhance various downstream tasks, including MoA prediction.[34] Given the field's rapid expansion in 2023-24, a detailed examination of representation learning strategies exceeds the scope as we confined our systematic review to articles published prior to the June 2023 cut-off date, making select exceptions for some articles published during to the time of writing. There is an ever-growing scope of representation learning in Cell Painting with the availability of large datasets.

**5.2 Cell Painting in Hit Discovery**

Cell Painting can also be used to potential 'hits'—compounds showing targeted biological activity—and to identify novel active compounds with therapeutic potential. While supervised ML models learn from previously defined categories, an outlier test such as novelty detection (ND) instead recognizes 'novel' (unknown) patterns, thus predicting whether compounds exhibit *any* biological activity.[73] For this, they used Cell Painting features for 641 validated and highly selective pharmaceutically relevant inhibitors over 123 targets to train models and they further used two additional validation sets (one for compounds that affect cell cycle and another with staurosporines). They found that a ND algorithm with an ensemble of classical ML algorithms was suitable to search for new active compounds. The ND method (LocalOutlierFactor from scikit-learn[39]) learns patterns of the data and compares the new samples with the learnt pattern by using a density function to determine whether new samples belong to the known dataset or not. Two NDs were trained; one with the bioactive compounds and another with the control groups and for a compound to be classified as "novel" both NDs should agree. Results on the two validation sets showed that for the cell cycle data, accuracy increased by 15% to 93% compared to not using ND, while specificity and precision remained the same and recall improved from 82% to 100%. For the other validation set, the performance of the model remained the same. This approach can be useful to identify bioactive compounds with 'novel' phenotypic responses. In addition, Nyffeler et al. compared various computational strategies to determine bioactivity hits using a Cell Painting assay and showed that nine out of ten approaches were highly concordant for 82% of the tested chemicals.[74] This indicates that Cell Painting assays contain a signal for bioactivity that can be used to predict assay hit calls via different approaches.

Cell Painting assays have also been used to elucidate the biological targets or actions of previously uncharacterized Dark Chemical Matter (DCM). These DCM compounds are usually analogs of bioactive compounds with drug-like features that lack biological activity in assays used to characterize the equivalent analogue structures. Pahl et al. profiled 7,700 DCM compounds with the Cell Painting assay and showed that 12% of them resulted in significant morphological changes (compared to a DMSO neutral control).[75] They selected 371 compounds and performed a cluster sub-profile analysis to identify their MoAs, comparing them with compounds with known MoAs (DNA synthesis inhibition, tubulin, and uncoupling of the mitochondrial proton gradient, among others). This analysis identified compounds associated with microtubule modulation, DNA synthesis, and pyrimidine biosynthesis.

**5.3 Cell Painting in Assay Activity Prediction**

Cell Painting data have proven useful in bioactivity prediction and drug-target interactions. For example, a landmark study by Simm et al. (2018) investigated using a three-channel glucocorticoid receptor (GCR) high-throughput imaging assay (in contrast to five-channel Cell Painting assays), which produced an 842-dimensional morphological fingerprint.[76] The multitask models developed in this study to predict assay activity performed highly (AUC>0.90) for a small subset, namely 31 out of 535 assays. Selecting two projects for prospective follow-up, they improved hit rates by 50-fold for a kinase target in an oncology project (from 0.725% to 36.3% hit rate) and by 289-fold for a non-kinase enzyme in a CNS project (from 0.088% to 25.5% hit rate). Both targets had no obvious connections to the glucocorticoid receptor for which the imaging assay was developed. This landmark study showed that these imaging assays generated broadly useful features that could then be used to predict activity on a wider range of biological targets.

Hofmarcher et al. then explored the ability of CNNs to predict the bioactivity of compounds in 209 biological assays.[77] They compared CNN models trained directly using Cell Painting images with fully connected neural networks (FNNs) trained using classically extracted numerical features and concluded that the former (mean AUC=0.73) were better than the latter (mean AUC=0.675). This highlighted that the raw-image CNN models could capture information, such as morphological characteristics, from Cell Painting images that may be overlooked by predefined CellProfiler features.[77] In addition to CNNs, transfer learning—where the knowledge of a pre-trained model (e.g., image-based knowledge) is transferred or fine-tuned to another model to perform a similar task (e.g. Cell Painting data)—is useful, reducing the need to train Deep Neural Network (DNN) models from scratch; this is being increasingly explored for cell imaging data especially when there is less data for the training of a DNN model.[78]

Various indicators of cell state or cell health can be detected using specific hypothesis-based dyes that reveal specific toxicity or anti-tumorigenic effects, such as apoptosis, DNA damage, and Reactive Oxygen Species (ROS) generation, among others. In contrast to the less expensive, hypothesis-free morphology signatures obtained in Cell Painting, indicators of cell state provide insight into exact cellular responses. That said, many aspects of cell health can be inferred from Cell Painting data (beyond a simple cell count), including the percentage of dead cells ($R^2$=0.62), number of S-phase cells ($R^2$=0.64), level of DNA damage in $G_1$-phase cells ($R^2$=0.51), and percentage of apoptotic cells ($R^2$=0.37).[30] Aside from providing insights into cell health, Cell Painting can also be used to study effects on individual sub-cellular compartments. For example, with multiple drugs where the endoplasmic reticulum (ER) was a downstream target, expectedly 80% of ER-related features were affected compared to other organelles.[79]

Other studies using similar ML methodologies compared bioactivity predictions based on Cell Painting and chemical structure information.[80,81] Prediction accuracies for targets like β-catenin (usually assayed using a specific stain) were better when using a BMF Macau model with Cell Painting profiles as side information (F1 score = 0.87) compared to a model using only chemical structural data (F1 score = 0.48).[80] Another study showed that the models that combined structural information and Cell Painting profiles, using similarities to training data, improved the AUC by 16.3% compared to models that only used chemical structure information.[81]

## 5.5 Phenotypic profiling of structurally diverse compounds

Considering structural diversity when creating small molecule libraries ensures the coverage of a vast spectrum of biological and functional relevance, facilitating the discovery of small molecules that can modulate specific targets. One way to generate a wide array of structurally distinct compounds is diversity-oriented synthesis (DOS).[82] These libraries could involve compounds that are structurally similar to known drugs or natural products. Phenotypic profiling, particularly with the Cell Painting assay, is increasingly being used to evaluate and characterize compounds generated through diversity-oriented synthesis. For example, in 2016, Nelson et al. explored the use of Cell Painting-based phenotypic profiling to compare the biological activity of certain $sp^3$-rich (carbon atoms with four single bonds) chemical compounds known as tetrahydrocyclopenta[c]pyranone derivatives.[83] They found that two of the epoxy ketone diastereomers synthesized caused striking cellular responses and induced consistent morphological changes for all doses, prompting studies to compare their morphological signatures to reference compounds. Studies using structurally diverse, reduced flavones and their Cell Painting profiles have shown that the fraction of $sp^3$ hybridized atoms is not the only factor for enhanced biodiversity, but stereochemistry and appendage diversity are also contributors.[84,85] More recently, biology-oriented synthesis has focused on pseudo-natural products, and to this end, Christoforow et al. characterized the potential bioactivity of novel classes of pyrano-furo-pyridone (PFP) pseudo-natural products.[86] They found that among the five initial hits (exhibiting bioactivity in the assay), the morphological profiles exhibited more than 70% similarity to the reference compound profiles; this then helped them to decipher their MoAs. Other studies have explored the use of target agnostic (hypothesis-free) Cell Painting to determine the phenotypic roles of novel compounds compared to reference compounds, including indocinchona alkaloids[87], a natural-product inspired flavonoid library[88]; spiroindane pyrrolidines[89]; pyrroquinoline pseudo-natural products[90]; and indofulvin pseudo-natural products[91]. Recently in 2023, a study by Liu et al. identified the mitotic kinesin, Eg5, as the molecular target for spirooxindoles—unique inhibitors of the kinesin Eg5 chemotype.[92] Overall, phenotypic profiling enables the bioactivity and MoA of diversity-oriented synthesized compounds to be predicted based on their similarities with reference compound activity profiles, which in many cases, can reveal new chemical scaffolds of interest.

**5.6 Predicting Compounds Toxicity**

More recently, Cell Painting assays have been used to predict multiple safety/toxicity-related assay outcomes, including effects on the cell cycle, cytokinesis, and cytoskeletal morphology. The Cell Painting assay was found sensitive enough to detect cell growth and viability at sub-lethal concentrations of podophyllotoxin, microtubule destabilizer, which resulted in a 50% reduction in cell count, even at the lowest tested concentration of 0.3 nM. [93] In addition, Seal et al. predicted the outcomes of 12 cytotoxicity- and proliferation-related *in vitro* assays using Cell Painting data profiles.[94] These models achieved an AUC of 0.71 compared with an AUC of 0.56 achieved by models using only chemical data (when using Morgan fingerprints). Using Cell Painting data to predict *in vitro* activity is not only limited to cell-based toxicity. Trapotsi et al. successfully predicted mitochondrial toxicity with an AUC = 0.93 when using Cell Painting profiles.[95]. In addition, in the same study they included both small molecule compounds and PROteolysis Targeting Chimeras (PROTACs), which have garnered attention due to their unique bifunctional nature and potential ability to degrade 'undruggable' targets.[95] Trapotsi et al. showed that Cell Painting could be used to identify PROTAC phenotypic signatures, but that these did not necessarily correlate with the Cell Painting profiles of their individual components (i.e., the POI ligand and the E3 ligase ligand). This further highlights the advantages of using Cell Painting for safety assessment of new therapeutic modalities where, as opposed to small molecules, no decades of experience and best practice have been established.

## 5.7 Phenotypic profiling of compound mixtures

Cell Painting has also been used to profile compound mixtures. For example, Pierozan et al. examined the responses of human breast epithelial cells to low-concentration mixtures of perfluorooctanoic acid (PFOS) and perfluorooctane sulfonic acid (PFOA), two widely used industrial chemicals.[96] Through Cell Painting, they were able to demonstrate the synergistic effects of these compounds on cell proliferation, even at very low concentrations (500 pM). This illustrates that the sensitive multiparametric nature of the Cell Painting assay can reveal alterations in cell morphology and toxicity. In another study, Rietdijk et al. have explored using Cell Painting to profile the effects of combining different environmental chemicals on four cell lines; in one example, Bisphenol A (BPA) did not cause significant morphological changes to cells when screened on its own. However, it caused various synergistic effects in three out of four cell lines (U-2 OS, A549 and MCF7 cells) when combined with the cationic compound that is used as an antibacterial and antifungal surfactant called Cetyltrimethylammonium bromide and Dibutyltin dilaurate, which is widely used as industrial chemical, serving as an antifouling coating, and in pesticides and fungicides.[97] In conclusion, Cell Painting assay was shown to elucidate biological effects exhibited by various mixtures of compounds.

One key Cell Painting development has been the assembly of sets of reference compounds with well-characterized toxic effects. These references can be used as benchmarks against which to compare novel compounds. Recently a study by Dahlin et al. established a reference set of 218 prototypical cytotoxic and nuisance compounds in U2OS using a concentration-response format (0.6–20uM). The compounds associated with cellular injury produced distinct morphological clusters (e.g., a genotoxin cluster and a tubulin poisons cluster). In addition, nonspecific and suboptimal probes 'historical' KAT inhibitors (hKATIs) produced profound morphological profiles, and compounds associated with cell damage, such as genotoxins, produced robust phenotypes. Concentration was an important parameter of cellular injury-related phenotypes. For example, some compounds, when present at higher concentrations trigger cellular responses called "cytotoxicity burst"; these higher compound concentrations activated multiple stress responses within the cell rather than affecting a singular target. This dataset can be used to characterize cellular injuries.[31]

Cell Painting is an *in vitro* assay that captures changes in cell morphology, but several studies have explored its ability to predict *in vivo* compound perturbation effects. In 2020, Nyffeler et al. performed an in-vitro-to-in-vivo extrapolation (IVIVE) of *in vitro* potency estimates obtained through Cell Painting.[98] They used reverse dosimetry to calculate administered equivalent doses (AEDs) and compared them to effect values from *in vivo* mammalian toxicity studies. They observed that 68% of the Cell Painting-based AEDs were either similar to or more conservative than the *in vivo* studies.[98] This shows that Cell Painting screens might be used to derive AEDs similar to traditional IVIVE practices. In 2023, Nyffeler et al. determined a phenotype altering concentration of compounds and used IVIVE to determine the administered equivalent doses (AEDs) to predict human exposure.[99] For 18 out of 412 chemicals, *in vivo*-relevant toxicity was observed at the given exposure, that is, the AEDs overlapped with predicted human exposures.[99] On a larger scale, the U.S. Environmental Protection Agency (EPA) is working towards using transcriptomics and Cell Painting data in their risk assessments.[100] In the future, new sources of relevant *in vivo* toxicity annotations, and the inclusion of pharmacokinetic (PK) information[101,102], might be used to augment Cell Painting data in predictive toxicology.

Overall, leveraging Cell Painting data has the potential to enhance toxicity predictions. Expanding the scope of the models using high-dimensional data can enable patterns and relationships to be identified effectively, offering the potential to triage compounds for certain hypothesis-based assays in future.

**5.8 Using Cell Painting Assays to Advance Disease Understanding**

Understanding disease biology and developing potential therapeutic interventions involves many steps, including disease modeling and biomarker discovery. The Cell Painting assay has been used to functionally associate human genes and disease-associated alleles based on the similar morphology of cells when those genes are perturbed, or alleles are present. Rohban et al. used this approach to reveal a previously unknown interaction between the NF-κB pathway and Hippo pathways, which regulate tumorigenesis and tumor progression[103], and furthermore to identify promising compounds that match a desired phenotypic profile impinging on those pathways.[104] Disease-specific morphological signatures can serve as biomarkers for disease diagnosis or prognosis, or to monitor therapeutic responses. For example, Cell Painting assays have been used to model cancer cell morphologies to identify the distinct morphological signatures associated with esophageal adenocarcinoma and responses to selective modulators for these phenotypes.[105,106] Cell Painting image-based data of cells overexpressing specific cancer-associated variants could also be used to predict the functional impacts of somatic variants, including at the single-cell level.[29]

Cell Painting has been used for antiviral drug discovery identifying virus-induced phenotypic signatures in virus-infected and non-infected cells.[26] In this study, they showed that treatment of infected cells with a panel of various host- and virus-targeting antivirals could reverse the morphological profile of the host cells towards that of a non-infected cell. It was also used to explore novel therapies for ER stress-associated disorders by identifying compounds that corrected aberrant morphological phenotypes associated with ER stress.[107] Cell Painting has also been used to investigate transcription factor EB (TFEB) signaling and lysosomal dysfunction by detecting phenotypic changes in organelles in response to TFEB localization.[108] Finally, the assay was also used to investigate drug resistance in anti-cancer therapy[109] by identifying the morphological signature of bortezomib treatment resistance in cells.

Another application of Cell Painting aimed to investigate relationships between genetic variants and cellular morphology in induced pluripotent stem cells (iPSCs).[110] Novel associations ("cell morphology quantitative trait loci" - cmQTLs) were identified between rare protein altering variants in WASF2, TSPAN15, and PRLR and morphological changes in the cell shape, nucleic granularity, and mitochondrial distribution. Further knockdown of these genes by CRISPRi confirmed their role in cell morphology, and hence morphological profiling can yield insight about the function of genes and variants. McDiarmid et al. previously used Cell Painting data to reveal 16 FDA-approved drugs from five mechanistic groups that were able to reverse morphological signatures associated with Alzheimer disease – risk gene SORL1 variants in neural progenitor cells.[111] Overall, the broad and multidimensional data generated by Cell Painting assays not only provides opportunities for new insights into complex cellular responses but can also reveal novel therapeutic targets and strategies for drug discovery and repurposing.

**5.9 Integrating Cell Painting, Transcriptomics, and Proteomics Data**

Given that Cell Painting readouts describe only one category of biological phenotype, one possibility to improve predictive models is to integrate Cell Painting data with further biological data, such as gene expression and proteomic data.[112] Nassiri and McCall integrated Cell Painting with LINCS (Library of Integrated Network-Based Cellular Signatures) gene expression data for improved insight into MoAs.[113] They used a reference database of 9,515 compounds to identify compounds with similar gene expression changes, followed by 'cell morphology enrichment analysis'. The enrichment analysis involved identifying significant associations between changes in cell morphology and gene expression, and then modeling these associations using ML and hence generating a dataset of the associations between genes and each morphological feature. They investigated the regulatory mechanisms underlying compound-induced changes in both gene expression and cell morphology for three compounds by performing pathway enrichment analysis. By integrating both Cell Painting and LINCS data and methods, this study revealed a novel interdependence between gene expression and cell morphologies and proposed using these relationships to infer compound MoAs.

Combining Cell Painting and gene expression data was also explored by Way et al., who showed that, together, they provided complementary information for mapping cell states.[44] Individually, Cell Painting could group compounds sharing the same MOA 44% of the time and mRNA profiles 50% of the time (across all doses), but when combined, attained 69% when comparing across doses.[44]

The benefits of integrating Cell Painting data also apply to protein profiling. Particularly, combining nELISA protein profiling and Cell Painting profiling data can provide complementary information. One study explored this by testing 306 well-characterized compounds with established MoAs using both profiling methods.[114] They then determined whether the two profiling methods were able to retrieve the known MoAs based on shared phenotypes. Cell Painting and nELISA profiling methods successfully retrieved compounds that shared at least one common MoA in 21.2% and 26.7% of cases, respectively. Some compounds were well predicted by both platforms while each platform showed better predictions for distinct subsets of compounds. Tian et al. found that combining morphology and chemical structure data significantly improved F1 detection scores for 10 well-represented MoA classes compared to using morphology data alone, with scores improving from 0.81 to 0.92, respectively.[115] Another study combined morphological profiling with proteome analyses to reveal lysosomotropic activity leading to cholesterol homeostasis disruption for the natural product-inspired compound, tetrahydroindolo[2,3-a]quinolizine derivative.[116]

Another study combined RNA-Seq and Cell Painting data to estimate the phenotype altering concentration of a set of 11 mechanistically diverse compounds, and found that for most of the compounds, the phenotype altering concentration from Cell Painting and biological phenotype altering concentration from RNA-Seq were within half an order of magnitude.[117] Furthermore, they found that combining both modalities provided the best potency estimates, particularly for compounds with strong morphological signatures that did not affect expression of target genes (*ATRA* in this study).

Morphological features and gene expression data were also used by Cerisier et al. to explore associations between chemicals and disease by developing a biological network combining chemical-gene-pathway-morphological perturbation and disease relationships.[118] They investigated two chemicals (amiodarone and prochlorperazine) because both showed a risk for drug induced liver injury (DILI) in humans and thus, they assessed if they share common information in Cell Painting and L1000 dataset. They found a direct relation between deregulated genes and cell morphology observations. This was one of the examples that the authors used to demonstrate that some compounds shared similar genes, pathways, and morphological profiles. Previously, combining proxy-DILI labels with chemical and pharmacokinetic features, achieving improved detection accuracy and differentiation between animal and human DILI sensitivity and it remains to be seen if -omics datasets such as Cell Painting, gene expression and proteomics data can be used for DILI prediction, which is one of the aims of the recently established OASIS consortium.[119,120] Overall, where multiple modalities can be considered together, that could help in elucidating chemical-phenotype observations.

In another study, Seal et al. combined gene expression and Cell Painting data to predict mitochondrial toxicity using machine learning models.[69] Their model, which incorporated information on gene expression, cell morphology, and chemical structure, improved F1 detection scores to 0.40 (from 0.25 using chemical structures alone). This was explored further with the recently developed similarity-based merger model, which increases the applicability domain as well as the diversity of assays that would be well predicted.[81] Similarity-based merger models combine the outputs of individual models trained on Cell Painting data and chemical structure information based on the structural and morphological similarities of the compounds in the test dataset to compounds in the training dataset. Models were trained for 177 assays to predict assay hit calls and the similarity-based merger models outperformed other models by an additional 20% assays (79 out of 177 assays) having an AUC > 0.70 compared to 65 out of 177 assays using chemical structure information alone and 50 out of 177 assays using Cell Painting information alone.

Most recently, Sanchez-Fernandez et al. developed CLOOME, a multi-modal contrastive learning algorithm, to combine chemical structure data and Cell Painting images into a unified space. Their retrieval system correctly identified the image corresponding to a given compound with an accuracy approximately 70 times higher than a random baseline model; this system was also used to predict compound activity (in a similar setting as Hofmarcher et al.) and CLOOME achieved an AUC of 0.714±0.20 across all prediction tasks.[121] This result indicates that the learned representations are transferable to different tasks (in this case bioactivity prediction) since no activity data were used to train the CLOOME encoders. Using images directly therefore enables unbiased insight into information contained within that image without requiring classical feature extraction algorithms. Overall, the integration of Cell Painting with diverse and complementary -omics modalities such as transcriptomics and proteomics can offer more comprehensive insights into the biological impacts of compounds and improve the accuracy of MoA predictions.

## CHALLENGES AND FUTURE DIRECTIONS

There are many avenues for improving image-based profiling in the future. One major challenge is the interpretation of morphological profiles. While sophisticated image analysis algorithms and machine learning methods can extract and analyze complex morphological signatures, interpreting these computational/statistical signatures can be challenging, even for classical algorithms where features are precisely defined mathematically. The BioMorph space attempts to address this by linking 827 Cell Painting features to 412 descriptive terms, based on mapping to assays capturing phenotypes of cell health[122]. However, for broad utility, mapping to more assay data would be needed.

A second challenge relates to data handling and storage.[123] The high-content nature of the Cell Painting assay generates vast amounts of data, which can be challenging to store, manage, and share, and the data requires considerable computational resource to process and analyze. Cloud-based solutions and open-source software tools are emerging to address these challenges[124], but increasing their user-friendliness would expand the use of this data type.

Another challenge is relevant to larger Cell Painting studies and common to all high-throughput assays: the adoption of assay miniaturization in academic and small company settings, to enable higher throughput and more cost-effective screening. 384-well format is commonly used for Cell Painting but requires high-throughput equipment as does the 1536-well format microplate recently demonstrated for Cell Painting assays in the industry.[125,126]

We see great promise in extending the Cell Painting assay (originally developed with 2D cell cultures) for use with more physiologically relevant systems such as 3D cell cultures, organoids[127], tissue slices, and live cell imaging.[128]

Finally, improvements in deep learning methods have dramatically altered the landscape in many fields of scientific research and we expect the same for Cell Painting data. Already, promising improvements have been made but there remains much room for improvement, particularly in batch correction methods that can extract biologically meaningful signals from technical noise.[45]

To conclude, over the past decade, the Cell Painting assay has evolved from a promising concept to a widely used tool for drug discovery and cellular biology. It has provided insights into the complex world of cellular morphology, expanding our understanding of disease mechanisms, and enhancing drug discovery processes. Continued advancements in Cell Painting (and advances towards extracting similar information from brightfield images) offer promising prospects, particularly for characterizing cellular responses, developing personalized medicine strategies, and providing deeper insights into complex biological processes.

**SIGNIFICANCE**

The adoption of Cell Painting and related technologies within the pharmaceutical industry and academic has been steadily growing. The availability of large Cell Painting datasets has empowered the exploration of machine learning methods on a wide range of biological endpoints. This expanded scope opens new possibilities for predictive modeling and comprehensive assessments of compound toxicity. However, transitioning from the introduction of a novel assay to achieving widespread implementation is not without its challenges, as recently noted for both image-based and transcriptional profiling.[129] It takes time to fund and conducting large-scale experiments, and accumulate experience. As observed with Connectivity Map—a genomic database released in 2006 that links drug compounds to gene expression profiles—launching a new tool or technique requires considerable time and evaluation under different contexts before it is widely adopted. Consortia can serve a valuable role in evaluation, offering several advantages, including pooling resources to increase sample sizes and the potential for experimental design to encompass a broader range of applications and chemical space. The involvement of multiple companies, organizations, and societies can foster collaborations and ensure the success of these endeavors. In summary, with the availability of larger datasets, increased industry interest, and the potential for collaboration through consortia, the future looks promising.

**METHODS**

**Literature Search**

The literature search was conducted using three major databases: ScienceDirect[20], PubMed[18], and Scopus[19], and completed in June 2023. All searches required the words "Cell Painting" to be present in the title, abstract, and subject terms/keyword headings. While writing this systematic review, we also searched the literature manually, which resulted in the addition of some articles published after June 2023 (listed in Supplementary Tables S1 and S2 released at https://github.com/srijitseal/CellPainting_SystematicReview).

**Inclusion and exclusion criteria**

Studies that used Cell Painting assays were included. Studies were excluded when they were not primarily in English, had been published before 2013 (before the Cell Painting assay was formally introduced), and had not been peer-reviewed (except in a few cases where we felt they were significant or included them during a manual search). Reviews, news articles, posters, thesis abstracts, and perspective papers were not included as primary research articles (these are instead listed in Supplementary Table S2 and referenced where applicable).

**Supplementary information**

The Supplementary Datasets are available as Supplementary Table S1: 89 studies included in this study (XLSX); Supplementary Table S2: 65 studies excluded from this study (XLSX) at https://github.com/srijitseal/CellPainting_SystematicReview

**Data availability**

No new datasets were used in this study.

**Code availability**

No code was used in this study.

**Author Contributions**

S. Seal and M.A. Trapotsi designed and performed the systematic review on studies using Cell Painting data. S. Seal and M.A. Trapotsi wrote the manuscript with extensive discussions with all authors. All of the authors reviewed, edited, and contributed to discussions on the manuscript and approved the final version of the manuscript.

**Funding**

S. Seal acknowledges funding from the Cambridge Centre for Data-Driven Discovery (C2D3) and Accelerate Programme for Scientific Discovery. A.E.C., S. Singh, and S. Seal acknowledge funding from the National Institutes of Health (R35 GM122547 to A.E.C.). O.S. acknowledges funding from the Swedish Research Council (Grants 2020-03731 and 2020-01865), FORMAS (Grant 2022-00940), Swedish Cancer Foundation (22 2412 Pj 03 H), and Horizon Europe (Grant Agreements 101057014 (PARC) and 101057442 (REMEDI4ALL)).

**Notes**

The authors declare the following competing financial interest(s): S. Singh and A.E.C. serve as scientific advisors for companies that use image-based profiling and Cell Painting (A.E.C.: Recursion, SyzOnc, Quiver Bioscience; S. Singh: Waypoint Bio, Dewpoint Therapeutics, DeepCell) and receive honoraria for occasional talks at pharmaceutical and biotechnology companies. J.C.P. and O.S. declare ownership in Phenaros Pharmaceuticals.

**Acknowledgments**

Figure 1 was created using BioRender. Figure 2-5 were created with DALLEv2 (https://openai.com/dall-e-2), Microsoft Designer (https://designer.microsoft.com/), and Bioicons (https://bioicons.com) which compiled images from the Database Centre for the life sciences/TogoTV (https://togotv.dbcls.jp) and Servier (https://smart.servier.com).